\RequirePackage{fix-cm}
\documentclass[smallextended]{svjour3}       
\smartqed  
\usepackage{graphicx}
\usepackage{lineno,hyperref}
\modulolinenumbers[5]

\usepackage{url}
\usepackage{amsmath}
\usepackage{amsfonts}
\usepackage{multirow}
\usepackage[section]{placeins}
\usepackage{algorithm} 
\usepackage{algpseudocode}
\usepackage{setspace}
\usepackage[table]{xcolor}
\usepackage{hhline}
\usepackage{subfig}
\usepackage{soul}
\usepackage{graphicx}
\usepackage{booktabs}
\usepackage{xspace}
\usepackage{amsmath}
\usepackage{xcolor}
\usepackage[misc]{ifsym}
\newcommand{\yh}[1]{\textcolor{black}{#1}}

\begin{document}

\title{ Sentiment Analysis and Topic Modeling for COVID-19 Vaccine Discussions}
 


\author{
        Hui Yin \and
        Xiangyu Song \and
        Shuiqiao Yang \and
        Jianxin Li*
}

\institute{\Letter Jianxin Li (the corresponding author), Hui Yin, and Xiangyu Song \at
              School of IT, Deakin University, Geelong, Australia \\
              \email{\{jianxin.li, yinhui, xiangyu.song\}@deakin.edu.au}           
           \and
           Shuiqiao Yang \at
             School of Computer Science and Engineering, University of New South Wales, Sydney, NSW 2032, Australia \\ 
              \email{shuiqiao.yang@unsw.edu.au}
}

\date{Received: date / Accepted: date}
\maketitle

\begin{abstract}
The outbreak of the novel Coronavirus Disease 2019 (COVID-19) has lasted for nearly two years and caused unprecedented impacts on people’s daily life around the world. 
Even worse, the emergence of the COVID-19 Delta variant once again puts the world in danger.
Fortunately, many countries  and companies have started to develop coronavirus vaccines since the beginning of this disaster. 
Till now, more than 20 vaccines have been approved by the World Health Organization (WHO), bringing light to people besieged by the pandemic.
The promotion of COVID-19 vaccination around the world also brings a lot of discussions on social media about different aspects of vaccines, such as efficacy and security.
However, there does not exist much research work to systematically analyze public opinion towards COVID-19 vaccines.
In this study, we conduct an in-depth analysis of tweets related to the coronavirus vaccine on Twitter to understand the trending topics and their corresponding sentimental polarities regarding the country and vaccine levels.
The results show that a majority of people are confident in the effectiveness of vaccines and are willing to get vaccinated.
In contrast, the negative tweets are often associated with the complaints of vaccine shortages, side effects after injections and possible death after being vaccinated.
Overall, this study exploits popular NLP and topic modeling methods to mine people's opinions on the COVID-19 vaccines on social media and to analyse and visualise them objectively.
Our findings can improve the readability of the noisy information on social media and provide effective data support for the government and policy makers.
\end{abstract}

\section{Introduction}
\label{sec_introduction}
The novel Coronavirus Disease (COVID-19) outbreak has taken an incalculable toll on the world, affecting the normal lives of billions of people worldwide. 
According to the latest news from the World Health Organisation, more than 200 million people have been infected, and over 4.42 million people have died of the COVID-19\footnote{https://covid19.who.int/}.
The impact of the COVID-19 pandemic is huge.
It is considered one of the most severe epidemics of this century, comparable to past pandemics such as the Spanish flu in 1918 and the Black Death in the mid-13th century \cite{zhou2021examination}.
Eliminating the COVID-19 has become the common goal for all countries.
Many countries and regions have adopted a series of specific measures to help slow down the spread of COVID-19.
Such as closing borders, reducing the activities in public places (e.g., restaurants, gyms, shopping centers), working/studying from home, restricting travel distance and maintaining good hygiene.
These measures have achieved good results in controlling the spread of the pandemic, and some restrictions have been lifted within some countries and cities. 
However, the recent COVID-19 Delta variant is more contagious, and countries worldwide have fallen into a state of emergency again.
It seems that the existence of COVID-19 will last for a long time in the future and vaccinations are the only long-term solution to the COVID-19 pandemic, provided that a majority of the population gets injections.
Thus, since the pandemic outbreak, countries and companies worldwide have started vaccine development and clinical trials.

As of August 23, 2021, there are 139 vaccine candidates and 22 of them have been approved in different countries across the world\footnote{https://covid19.trackvaccines.org/}.
The Pfizer/BioNTech vaccine was the first to receive emergency validation by the World Health Organization (WHO) on December 31, 2020\footnote{https://www.who.int/news/item/31-12-2020-who-issues-its-first-emergency-use-validation-for-a-covid-19-vaccine-and-emphasizes-need-for-equitable-global-access}, followed by AstraZeneca, Covishield, Janssen, Moderna.
Russia has become the first country in the world to grant regulatory approval to a COVID-19 vaccine after less than two months of human testing\footnote{https://www.abc.net.au/news/2020-08-11/russia-approves-first-coronavirus-vaccine-vladimir-putin-says/12547608}.
Subsequently, each country approved a number of vaccines and formulated particular policies to encourage all citizens to get vaccinated.
According to statistics, 31.7\% of the world population have received at least one dose of a COVID-19 vaccine, and 23.7\% are fully vaccinated\footnote{https://ourworldindata.org/covid-vaccinations}.
The rate of vaccinated people against COVID-19 (fully and partly vaccinated) is 59.47\% in the United States, 72.69\% in Canada, and 69.73\% in the United Kingdom. 
In contrast, some countries have very low vaccination rates, for example, 31.06\% in India, 9.22\% in Iran, 42.29\% in Mexico, and 18.09\% in Pakistan.
The current vaccination rate has not yet reached the minimum requirement for controlling the spread of the pandemic in countries.
One possible reason is that people do not know enough about the vaccine, lack confidence in it, and are skeptical about its safety.
They fear that the vaccine may cause long-term chronic illness because the vaccines have not been tested sufficiently.
Another possible reason is that the spread of fake information related to COVID-19 on social media may encourage those who are hesitant or doubtful about the vaccine to go against it.
For example, there is a conspiracy theory that denies the existence of the coronavirus epidemic. Therefore, it is necessary to figure out people's concerns about vaccines in the promotion of vaccines.

Social media (e.g., Twitter, Facebook, Instagram) and online forums (e.g., StackOverflow, Kaggle, Yahoo) provide a convenient and trustable information source for researchers \cite{alduaiji2018influence,li2020community,yin2020deep,yang2017modeling}.
People can freely post, comment, express their opinions on specific topics or communicate with others on these platforms \cite{yang2019discovering,jiang2019sentence}.
Therefore, the discussion of the COVID-19 vaccine on social media provides us with a source of data to find out people's concerns about the vaccine.

\begin{figure}[!htb] 
\centering
\includegraphics[width=100mm, scale=1]{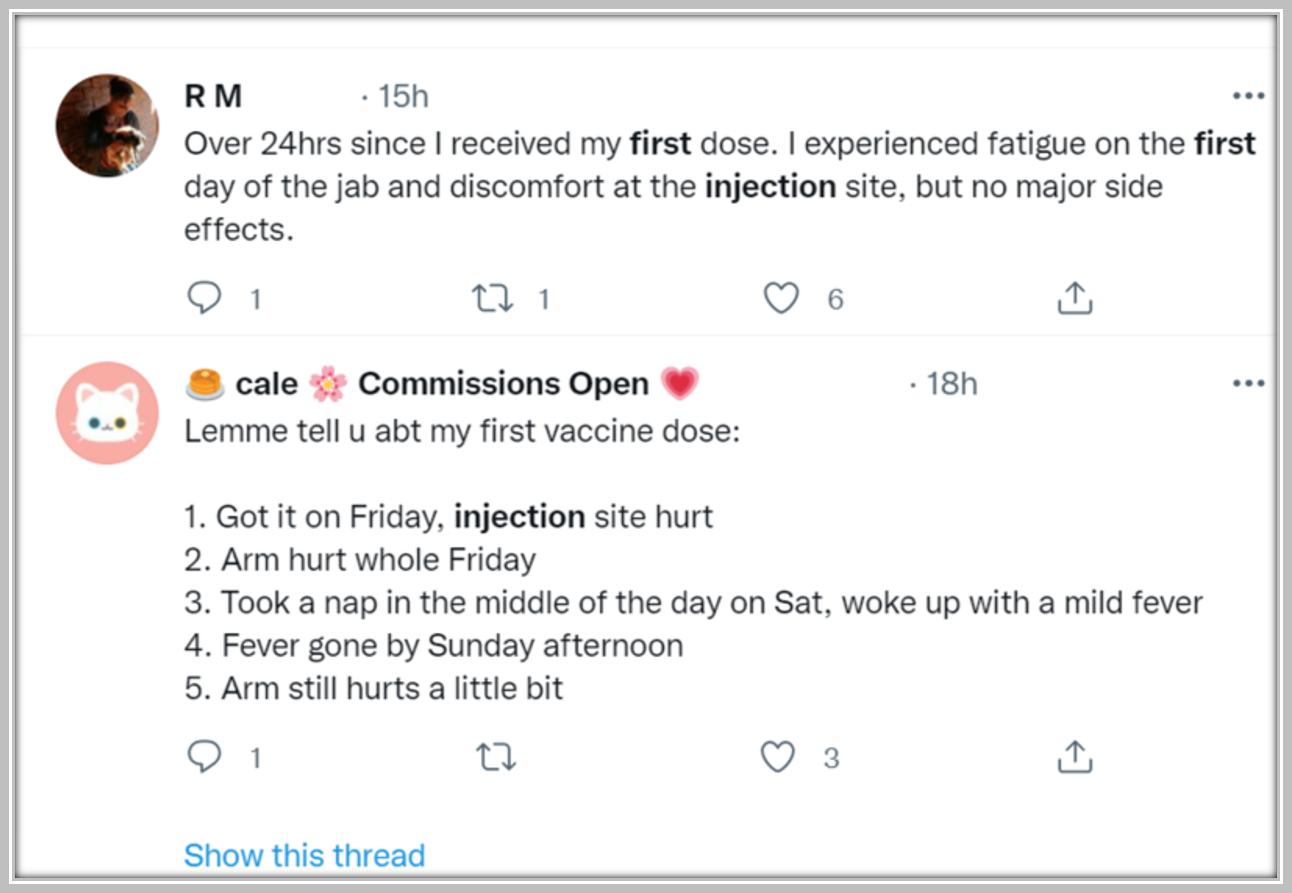}
\caption{Example tweets about COVID-19 vaccine discussions on Twitter.} 
\label{fig_TweetsExample}
\end{figure}

\yh{This paper examines the tweets about the coronavirus vaccine on Twitter,} extract the topics of concern, and the sentiment polarity in the tweets.
Rather than looking at the whole world, this paper examines four countries with the highest number of tweets during the study period.
These tweets dominate the direction of public opinion and usually represent the majority of people's opinions.
In addition, analyzing people's attitudes towards different vaccines and performing visual analyses can help the government understand the people's conditions and take corresponding measures more objectively if necessary.

\yh{The highlights for this work are summarized as follows:
\begin{itemize}
    \item To the best of our knowledge, this is the first analysis of the public discussions related to the COVID-19 vaccines on social media since the emergence of the COVID-19 Delta variant.
    \item We adopted two robust text mining techniques: Latent Dirichlet Allocation (LDA) and Valence Aware Dictionary and Sentiment Reasoner (VADER), to extract the hidden information buried in noisy social media discussions.
    \item We conducted a comprehensive analysis of COVID-19 related discussions on Twitter. We found that discussions are mostly positive, and the dominant sentiment of trust means a high acceptance of the COVID-19 vaccine.
\end{itemize}
 }
\yh{The paper is organised as follows. We  review the related work in Section \ref{sec_relatedwork}.
Section \ref{sec_dataset} introduces the process of data preprocessing and makes an in-depth exploration of the dataset}.
The methods used in this study are detailed in Section \ref{sec_methods}.
Section \ref{sec_results} introduces the process of data analysis and visualizes the results. 
Conclusions are made in Section \ref{sec_conclusions}.
\section{Related Work}
\label{sec_relatedwork}
\yh{Social media data analysis has been widely used for health-related issues and emerging public health crises \cite{du2019neural,sarki2020automated}.
Since the outbreak of the COVID-19 pandemic, there have been many discussions on social media every minute, such as Twitter, Facebook, Weibo.
Such a large number of posts provide a valuable source of data and has received attention from researchers \cite{wang2020efficient,tian2019evidence,yin2021vulnerability,zhang2020decision}.
Some research has been carried out on tweets related to the COVID-19 pandemic, analyzing and mining helpful information hidden in the posts.}
\subsection{Sentiment Analysis for COVID-19 based on Twitter}
Some work exploited sentiment analysis as a tool to investigate people’s reactions during the pandemic through 
their posts on social media.
Li et al. \cite{li2020analyzing} analyzed the posts of Americans and Chinese on Twitter and Weibo during the pandemic from January 20, 2020 to May 11, 2020.
They compared the emotions (i.e., anger, disgust, fear, happiness, sadness, surprise) and the emotional triggers (e.g., what a user is angry/sad about) to reveal sharp differences in 
how people perceive COVID-19 in different cultures.
Stella et al. \cite{stella2020lockdown} investigated the emotional and social repercussions during the lockdown in Italy, the first country to respond to the threat of COVID-19 with a national lockdown. They discovered the emergence of complex emotions in which fear and anger coexisted with solidarity, trust, and hope.
Dubey \cite{dubey2020twitter} conducted text mining and sentiment analysis on COVID-19 tweets from 12 countries to analyze how people in different countries/regions responded to these situations.
The results showed that most people were confident in controlling the pandemic, but there was also fear, sadness and disgust around the world.
Zhou et al. \cite{zhou2021detecting} extracted five months of COVID-19 related tweets on Twitter to analyze the sentiment dynamics of people living in the state of New South Wales (NSW), Australia during the pandemic period.
They divided tweets according to local government areas (LGA) and observed the dynamic changes in sentiment over time.
Yin et al. \cite{yin2020detecting} proposed a novel framework to dynamically analyze the topic and sentiment of 13 million tweets related to COVID-19. They found that the proportion of positive tweets was higher than negative tweets during the study period (2 weeks), which is consistent with other similar work. 
This work further analyzed the daily hot topics about the COVID-19 pandemic and found the common concerns discussed by people during the study period.
For example, staying at home to ensure safety, the latest case reports, and people dying from the pandemic.
\subsection{Infodemic Analysis for COVID-19 based on Twitter}
In addition to being a valuable data source, social media has been described as a source of toxic ``infodemic” (i.e., information of questionable quality).  
During the COVID-19 pandemic, vast infodemics have been generated worldwide mixed with false/fake or misleading information in the digital and physical environment.
It causes confusion and risk-taking behaviors that can harm health, leads to mistrust in health authorities and undermines public health response\footnote{https://www.who.int/health-topics/infodemic}.
Some work has focused on this type of information on social media during the pandemic period.
Yang et al. \cite{yang2020analysis} comprehensively studied the spread of prevalent myths related to COVID-19, 
people's participation with them, and people's subjective feelings about myths.
They found that myths about the spread of infection and preventive measures spread faster than other myths, such as ``5g corona is truth'', ``Eating curry can prevent the COVID-19''.
People were most worried about the spread of coronavirus, and the common emotion among people was fear.
Gallotti et al. \cite{gallotti2020assessing} noticed that infodemic 
spread rapidly and widely through social media platforms during the pandemic.
This information may mislead the public or increase social panic.
Therefore, while the government and the people were fighting against the COVID-19 virus, they must also fight against infodemic.	
They analyzed more than 100 million Twitter messages posted worldwide during the early stages of the epidemic and then classified the reliability of the news being circulated.
Furthermore, an Infodemic Risk Index was developed to capture the magnitude of exposure to unreliable news across countries.
To contribute to the fight against the infodemic, Bang et al. \cite{bang2021model} aimed to achieve a robust model for the COVID-19 fake-news detection task proposed in CONSTRAINT 2021 (FakeNews-19). 
They further improved the robustness of the model by evaluating different COVID-19 misinformation test sets (Tweets-19) to further improve the generalization ability of the model to solve the COVID-19 fake news problem in online social media platforms.

\subsection{Analysis of COVID-19 Vaccine Discussion on Social Media} 
With the development and promotion of vaccines, many researchers have carried out research work on COVID-19 vaccine related discussions on social media.
Kwok et al. \cite{kwok2021tweet} extracted topics and sentiments related to the COVID-19 vaccine from Australian Twitter users between January and October 2020 .
They employed R library package \textit{syuzhet} to score each tweet into two sentiments (positive, negative) and eight emotions (anger, fear, anticipation, trust, surprise, sadness, joy, and disgust).
They found that two-thirds of all tweets expressed positive opinions and one-third expressed negative opinions.
Finally, they identified 3 LDA topics in the dataset: (1) attitudes toward COVID-19 and its vaccination, (2) advocacy of infection control measures against COVID-19, and (3) misconceptions and complaints about COVID-19 control.
Lyu et al. \cite{lyu2021covid} used the same methods as \cite{kwok2021tweet} to identify sentiments and topics over a long time span in public discussions related to the COVID-19 vaccine on social media, with the goal of better understanding public perceptions, concerns and emotions that may influence the achievement of herd immunity goals.
For the topic modeling, they yielded 16 topics, which were grouped into five overarching themes.
Bonnevie et al. \cite{bonnevie2021quantifying} quantified the increase in Twitter conversations around vaccine opposition during the COVID-19 pandemic in the United States.
They first collected such tweets, classified them into topics, and then tracked them.
After four months of observation, they found a noticeable increase in vaccine opposition on Twitter.
Exposure to these increased amounts of vaccine opposition may mislead people to oppose vaccines, which could have a drastic impact on the health of populations for decades to come.
Therefore, to ensure the widest support for a COVID-19 vaccine, it is essential to identify and address the messages used by vaccine opponents. 
Thelwall et al. \cite{thelwall2021covid} conducted a study to understand what types of vaccine hesitancy information shared on Twitter might be helpful in designing interventions to address misleading attitudes.
The main themes discussed were conspiracies, vaccine development speed, and vaccine safety.
The majority (79\%) of those who refused vaccines on Twitter expressed right-wing views, fear of the deep state, or conspiracy theories. A significant proportion of those who refused vaccination (18\%) tweeted about other topics in a mainly apolitical manner.

\section{\yh{Data Preprocessing and Statistics}}
\label{sec_dataset}
For this study, we focus on analyzing the topics and sentiments of the COVID-19 vaccine related tweets on Twitter.
As of August 23, 2021, there are 139 vaccine candidates, and 22 of which have been approved by different countries, and 192 countries with approved vaccines.
For example, vaccines such as Pfizer, Oxford/AstraZeneca and Sinovac have been approved in the United States, the United Kingdom and India, respectively. 
We adopt the latest publicly available dataset of the COVID-19 vaccine tweets from Kaggle\footnote{https://www.kaggle.com/gpreda/all-covid19-vaccines-tweets}.
The period for the collected tweets is from December 12, 2020 to July 2, 2021, and the dataset covers seven popular vaccine brands\footnote{https://covid19.trackvaccines.org/vaccines/approved/\#vaccine-list}, as shown in Table~\ref{tab_VaccineBrand}:
\begin{table}[H]
\centering
\caption{The COVID-19 vaccine brands in the dataset.}
\renewcommand{\arraystretch}{1.4}
\begin{tabular}{l|l}
\toprule
\toprule
\textbf{Vaccine Brand}                 & \textbf{Description}                              \\ \midrule
\textbf{Pfizer/BioNTech} & Approved in 97 countries, 27 trials in 15 countries.          \\ \hline
\textbf{Sinopharm} & Approved in 60 countries, 9 trials in 7 countries.                         \\ \hline
\textbf{Sinovac} & Approved in 39 countries, 19 trials in 7 countries.                      \\ \hline
\textbf{Oxford/AstraZeneca} & Approved in 121 countries, 39 trials in 20 countries. \\ \hline
\textbf{Moderna} & Approved in 69 countries, 25 trials in 6 countries. \\ \hline
\textbf{Covaxin} & Approved in 9 countries, 7 trials in 1 countries.             \\ \hline
\textbf{Sputnik V} & Approved in 71 countries, 20 trials in 7 countries.\\  \bottomrule \bottomrule
\end{tabular}
\label{tab_VaccineBrand}
\end{table}
We preprocessed the original dataset by the following steps.
Firstly, as the location of a tweet is necessary information in this study, we first deleted the tweets without location information and got 78,827 tweets.
After that, we removed the noisy words from the remaining tweets. The procedures include: (1) Removing the Twitter handles, URLs, emojis, and hashtags; (2) Removing non-English words or common words that do not provide insights into a specific topic (e.g., stop words);
(3) Case folding (i.e., lowering the case of words to allow for lexical processing);
(4) Lemmatization to remove inflected endings and return a word to its base or dictionary form. 
(5) Investigating the combination of two words (bigrams) to ensure that words such as ``side\_effect” could be one token instead of separating ``side” and ``effect”.
We also removed tweets with a length less than four after processing, which usually cannot provide reasonable semantics.
In the end, we got 75,665 tweets for our experimental study.

\begin{figure}[h]
\centering
\includegraphics[width=80mm, scale=0.8]{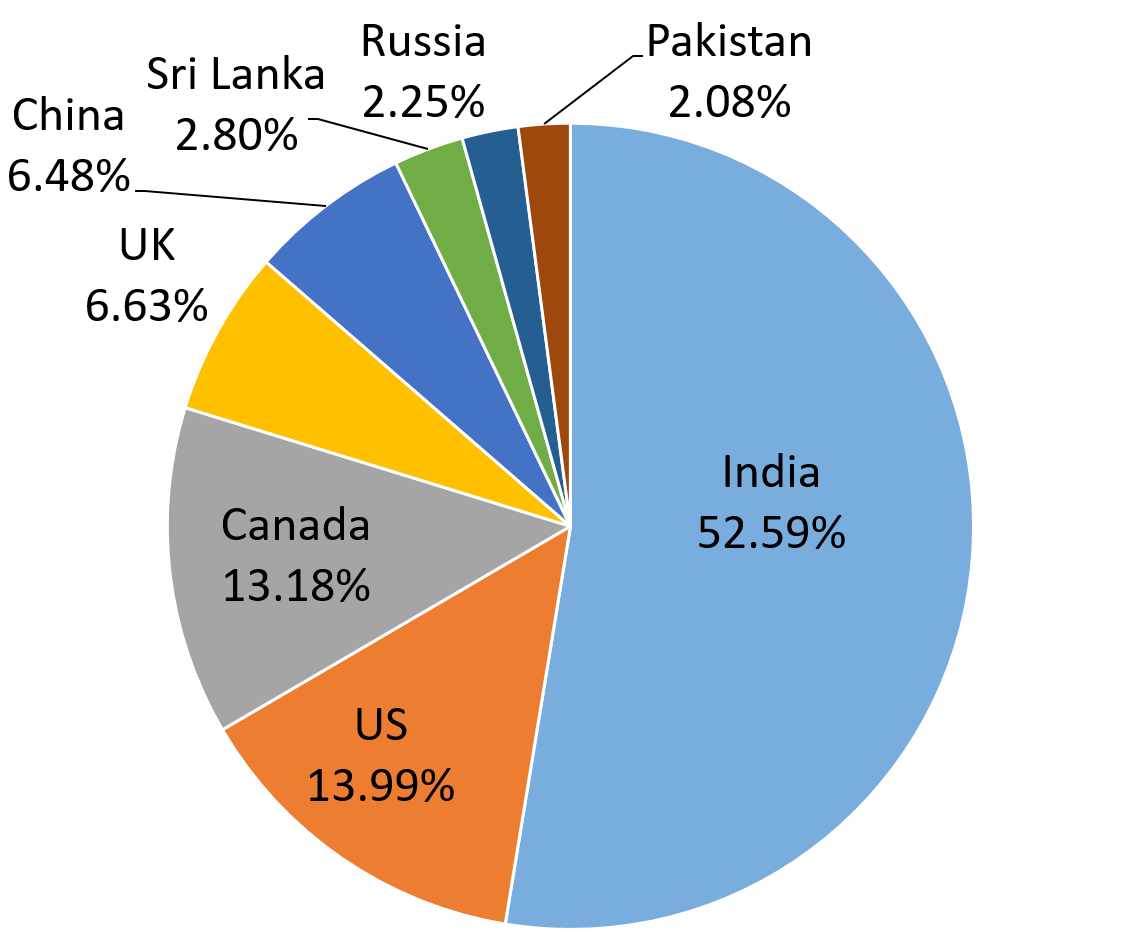}
\caption{The distribution of tweets of top eight countries in the dataset.} 
\label{fig_TweetsPerCountry}
\end{figure}

Fig.~\ref{fig_TweetsPerCountry} shows the top 8 countries with the largest number of tweets and their proportions.
India accounts for more than half of the posts, which is 52.59\%.
Such a high volume of tweets may be related to the out-of-control pandemic in India and a large number of users.
People are more actively participating in the discussion of the COVID-19 vaccine.

According to the COVID-19 vaccine official website\footnote{https://covid19.trackvaccines.org/trials-vaccines-by-country/}, in Table~\ref{tab_ApprovedVaccine}, we list the approved vaccines for the four countries with the most tweets in the dataset.
In fact, the vaccines approved in each country are not limited to these seven brands, but we only study the most popular seven vaccines in this study.

\begin{table}[!htbp]
\centering
\caption{Statistics on COVID-19 vaccines approved in four countries. We only count the seven brands included in the dataset. In fact, every country has approved more vaccines.}\label{tab_ApprovedVaccine}
\renewcommand{\arraystretch}{1.4}
\begin{tabular}{l|cccc}
\toprule
\toprule
                                                          \textbf{Vaccine Brand}                        & \multicolumn{1}{l}{\textbf{India}} & \multicolumn{1}{l}{\textbf{US}} & \multicolumn{1}{l}{\textbf{Canada}} & \multicolumn{1}{l}{\textbf{UK}} \\ \midrule 
 \textbf{Pfizer/BioNTech} & {}                          & {$\surd$}                           & {$\surd$}                          & {$\surd$}      \\         
\textbf{Sinopharm}                                                                &                                    &                                      &                                     &    \\                    
\textbf{Sinovac}                                                                  &                                    &                                      &                                     &      \\                    
\textbf{Oxford/AstraZeneca}                                                       & $\surd$                                  &                                      & $\surd$                                   & $\surd$               \\ 
\textbf{Moderna}                                                       & $\surd$                                  & $\surd$                                    & $\surd$                                   & $\surd$    \\                  
\textbf{Covaxin}                                                  & $\surd$                                  &                                      &                                     &      \\                           
\textbf{Sputnik V}                                                      & $\surd$                                  &                                      &                                     &                             
\\ \bottomrule \bottomrule
\end{tabular}
\end{table}

\section{\yh{Adopted Methods for Topic Modeling and Sentiment Analysis}}

\label{sec_methods}
We employ two methods for in-depth analysis of the COVID-19 vaccine related tweets on Twitter.
The first one is Valence Aware Dictionary for sEntiment Reasoning (VADER) for sentiment analysis, and the second one is Latent Dirichlet Allocation (LDA) for topic modeling.

\subsection{Measuring Tweet Sentiment}
Sentiment analysis (SA), also known as opinion mining, aims to automatically mine the opinions, attitudes, and feelings in texts, has a wide range of applications.
We use VADER \cite{hutto2014vader} to analyze the sentiment polarity of tweets in this study.
VADER is a lexicon and rule-based sentiment analysis tool specifically attuned for sentiments expressed in social media.
One of its most significant advantages is that it can be used directly on the original tweet with no preprocessing of the text.
Some examples of VADER scoring results\footnote{https://github.com/cjhutto/vaderSentiment} are shown in Table~\ref{tab_VADERexample}. VADER provides three types of scores, which are positive, neutral and negative.
In addition, VADER can also give a compound score, which is very effective when dealing with social media data.

VADER uses the compound score as the final sentiment score of a sentence. 
The compound score is computed by summing the valence scores of each word in the lexicon, adjusted according to the rules, and then normalized to be between -1 (most extreme negative) and +1 (most extreme positive).
\begin{table}[h]
\caption{Some examples of VADER scoring results.}
\centering
\renewcommand{\arraystretch}{1.3}
\begin{tabular}{ l }
\toprule
\toprule
\textbf{Examples of tweets and VADER scoring}   \\ \midrule
VADER is VERY SMART, uber handsome, and FRIGGIN FUNNY!!! \\
\{'pos': 0.706, 'neu': 0.294, 'neg': 0.0, `compound': 0.9469\} \\ [4pt] \hline
Today SUX!\\
\{`pos': 0.0, `neu': 0.221, `neg': 0.779, `compound': -0.5461\} \\ [4pt] \hline
Make sure you :) or :D today! \\ 
\{`pos': 0.706, `neu': 0.294, `neg': 0.0, `compound': 0.8633\}  \\ [4pt] \hline
Catch utf-8 emoji such as :-D, :-) \\ 
\{`pos': 0.279, `neu': 0.721, `neg': 0.0, `compound': 0.7003\}   \\ [4pt] \hline
Not bad at all \\
\{`pos': 0.487, `neu': 0.513, `neg': 0.0, `compound': 0.431\}   \\ \bottomrule \bottomrule                                                  
\end{tabular}
\label{tab_VADERexample}
\end{table}
 
We set a standardized threshold for classifying sentences as positive, neutral, or negative, as follows:
\begin{equation}
    S_{f_i}=\begin{cases}positive & v_{score} >=0.05, \\negative & v_{score} <=-0.05, \\ neutral & otherwise,  \end{cases}
\end{equation}
where $v_{score}$ is the compound score of the $i$-th tweet,  $S_{f_i}$ is the final polarity of tweet.  
If the compound score $v_{score}$ is not less than 0.05, the sentence is considered to be positive. If the score is not greater than -0.05, its polarity is negative. Otherwise, the sentence polarity is neutral. 
Table~\ref{tab_TweetSAExample} shows examples of tweets with positive and negative sentiment scores computed with VADER in this study.

\begin{table*}[h]
\centering
\caption{Examples of tweets with positive/negative sentiment.}
\renewcommand{\arraystretch}{1.4}
\begin{tabular}{l|l}
\toprule
\toprule
\textbf{Tweets} & \textbf{Sentiment} \\  \midrule

\begin{tabular}[c]{@{}l@{}}thanks to the vaccines, i was able to give my grandma a hug \\today for the first time in a long time   \end{tabular} & Positive \\ \hline
\begin{tabular}[c]{@{}l@{}}got my 2nd  shot yesterday; my arm hurts a little more\\than after the 1st, but glad to be fully vaccinated. \end{tabular} & Positive \\ \hline
\begin{tabular}[c]{@{}l@{}}i received the first vaccine. thank you and i am grateful \end{tabular} & Positive \\ \hline
\begin{tabular}[c]{@{}l@{}}just received my second dose of  happy dance to commence. \end{tabular} & Positive \\ \hline
\begin{tabular}[c]{@{}l@{}}its been a month since my  dose number two and i am\\concerned that my shoulder might be permanently jacked up.\end{tabular} & Negative \\ \hline
\begin{tabular}[c]{@{}l@{}}i am lost for words with reports that people in the eu are\\refusing the  vaccine.\end{tabular} & Negative \\ \hline
\begin{tabular}[c]{@{}l@{}}my second dose of of  is due in 4 days and there is no stock\\or dates available.
what do i do now? \end{tabular} & Negative \\ \hline
\begin{tabular}[c]{@{}l@{}}11.5 hours later my arm hurts and the upper part is visibly \\ swollen and i can
feel a large lump.\end{tabular} & Negative \\  \bottomrule \bottomrule
\end{tabular}
\label{tab_TweetSAExample}
\end{table*}


\subsection{Topic modeling of Tweets}
\label{TopicmodelingOfTweets}
\begin{figure}[] 
\centering
\includegraphics[width=100mm, scale=1]{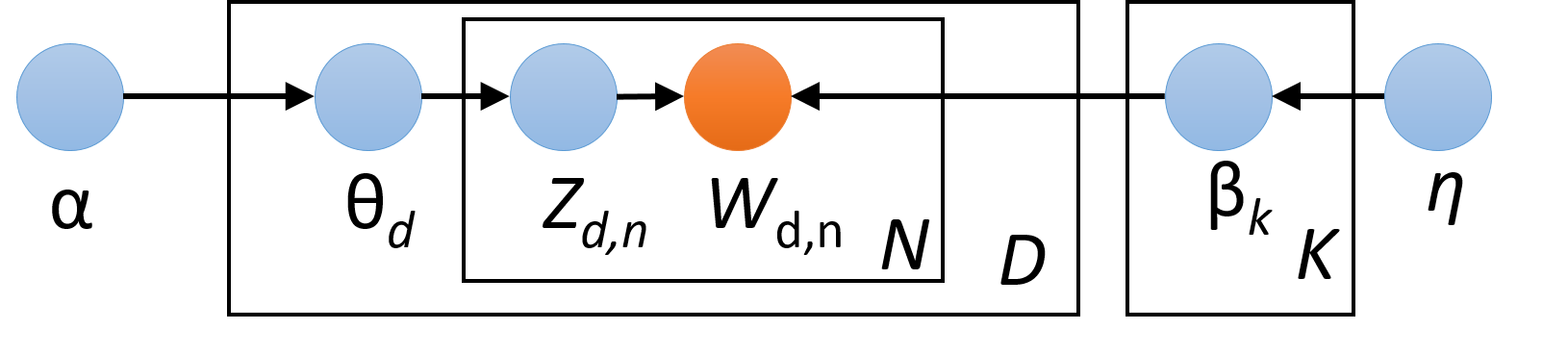}
\caption{A plate notation explanation of LDA.} 
\label{fig_LDA}
\end{figure}

Topic modeling is a method for the unsupervised classification of documents. Specifically, it's the process of learning, recognizing, and extracting high-level semantic topics across a corpus of unstructured text even when people are unsure what they are looking for.
It is a great way to get a bird's-eye view of a large text collection.
The most popular topic model is Latent Dirichlet Allocation (LDA) proposed by Blei et al. \cite{blei2003latent}, LDA aims to find topics a document belongs to, based on its words. 
LDA is based on a Bayesian probabilistic model where each topic has a discrete probability distribution of words, and each document is composed of a mixture of topics.
In LDA, the topic distribution is assumed to have a Dirichlet prior, giving a smooth topic distribution for each document.
The probability for a corpus is modeled in Eq. \ref{eq_LDA}, where the documents and words are assumed to be independent. 
We show the plate notation explanation of LDA in Fig.~\ref{fig_LDA} while the meaning of the notations is shown in Table~\ref{tab_NotationMeaning}. 

\begin{equation}
\begin{aligned}
 & \prod_{d=1}^{N_d}P({w_1},\cdot\cdot\cdot,{w_{N_d}}|\beta,\alpha)= 
&\prod_{d=1}^{N_d}\int_{\theta_d}P({\theta_d}|\alpha)\left\{\prod_{n=1}^{N_d}\left(\sum_k{\theta_{dk},\beta_{kw_n}}\right)\right\}d{\theta_d}
\end{aligned}
\label{eq_LDA}
\end{equation}

\begin{table}[H]
\centering
\caption{Meaning of the notations.}
\centering
\renewcommand{\arraystretch}{1.4}
\begin{tabular}{l|l}
\toprule
\toprule
\textbf{Symbol}                 & \textbf{Description}                              \\ \midrule
\textbf{$K$} & total number of topics          \\ \hline
\textbf{$D$} & total number of documents                         \\ \hline
\textbf{$N$} & total number of words in a document                      \\ \hline
\textbf{$\alpha,\eta$} & Dirichlet parameters                        \\ \hline
\textbf{$\theta_d$} & per-document topic proportions \\ \hline
\textbf{$Z_{d,n}$} & per-word topic assignment              \\ \hline
\textbf{$W_{d,n}$} & observed word                  \\ \hline
\textbf{$\beta_{k}$} & topic, a distribution over the vocabulary             \\ \bottomrule \bottomrule
\end{tabular}
\label{tab_NotationMeaning}
\end{table}

LDA assumes the following generative process for a corpus $D$ consisting of $M$ documents each of length $N_i$:
\begin{enumerate}
    \item Generate ${\theta_i}\sim Dir({\alpha})$, where $i \in\left\{1,2,\cdot\cdot\cdot,D\right\}$. $Dir({\alpha})$ is a Dirichlet distribution with symmetric parameter ${\alpha}$ where ${\alpha}$ is often sparse.
    \item Generate ${\beta_k} \sim Dir ({\eta})$, where $k \in\left\{1,2,\cdot\cdot\cdot,K\right\}$ and ${\beta}$ is typically sparse.
    \item For the $n_{th}$ position in document $d$, where $n \in\left\{1,2,\cdot\cdot\cdot,N_d\right\}$ and $d \in\left\{1,2,\cdot\cdot\cdot,D\right\}$.
    \begin{enumerate}
        \item Choose a topic $z_{d,n}$ for that position which is generated from $z_{d,n} \sim Multinomial ({\theta_i}$)
        \item Fill in that position with word $w_{d,n}$ which is generated from the word distribution of the topic picked in the previous step $w_{i,j} \sim Multinomial (\theta_{z_{d,n}})$
    \end{enumerate}
\end{enumerate}

\begin{figure}[] 
     \centering
     \includegraphics[width=80mm, scale=0.8]{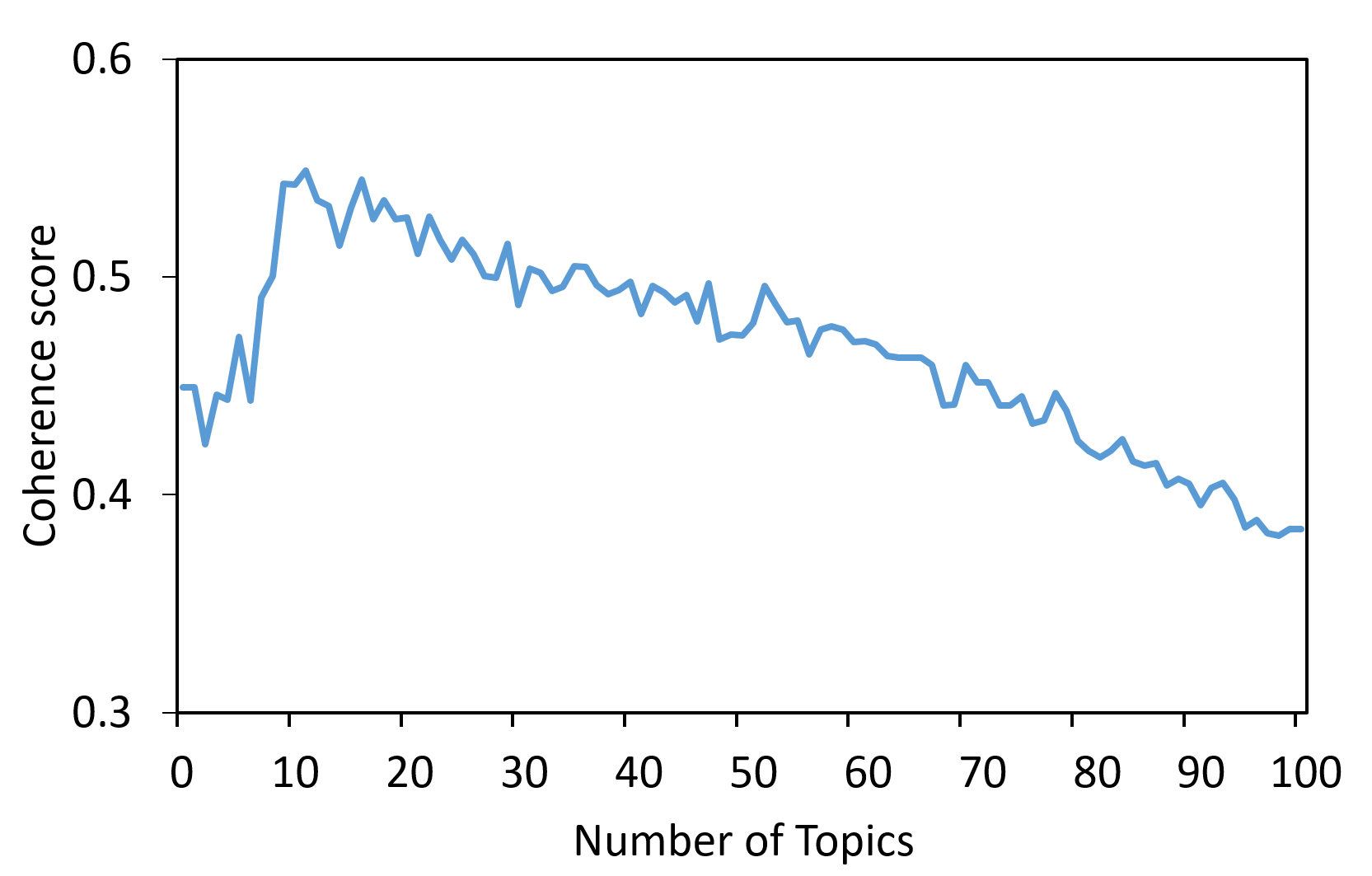}
    \caption{Coherence scores corresponding to the different number of topics.}
    \label{fig_CoherenceScore}
\end{figure}

In this study, we employ LDA for topic modeling and discuss hot topics in positive and negative tweets separately.
The number of topics is a crucial parameter in topic modeling. To make these topics human interpretable, we use the coherence score to determine the optimal number of topics.
The coherence score in the following Eq. \ref{coherence} helps to distinguish between human understandable topics and artifacts of statistical inference: 
\begin{equation}
\label{coherence}
    Coherence=\sum_{i<j}score(w_i,w_j). 
\end{equation}
The coherence selects top $n$ frequently occurring words in each topic, then aggregates all the pairwise scores of the top $n$ words $w_i,\cdot\cdot\cdot,w_n$ of the topic. 
Finally, we can get the total coherence score of the current number of topics. 
Fig.~\ref{fig_CoherenceScore} displays the coherence score of all tweets for the number of topics across two validation sets, and a fixed $\alpha = 0.01$ and $\beta = 0.1$.
We set the range of the number of topics from 1 to 100. According to the results, the coherence score is highest when the number of topics is 11, so we determine the number of topics to be 11, and then perform LDA topic modeling on the tweets.

\section{\yh{Experimental Results and Analysis}}
\label{sec_results}
We first look at the high-frequency vocabulary in the dataset, and then we extract prevalent words in tweets concerning user location and vaccine brands.
After that, we use VADER to generate the sentiment polarity of each tweet, namely positive, negative and neutral, and then further analyze the attitudes of users in various countries to the seven vaccines.
Finally, we use the LDA topic model to generate the topics of positive and negative tweets and examine the hot topics discussed in the tweets, respectively.

\subsection{Prevalent Words by Countries}
\begin{figure}[h]
\centering
\includegraphics[width=100mm, scale=1]{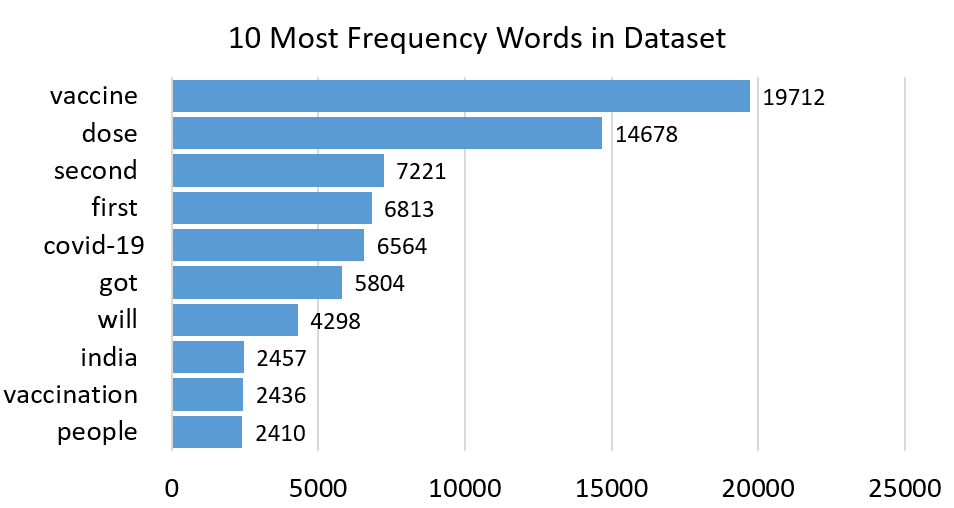}
\caption{The 10 most high-frequency words in the dataset.} 
\label{fig_MostFrequencyWords}
\end{figure}

After removing the stopwords and meaningless words, we first count the high-frequency vocabularies in the dataset, as shown in Fig.~\ref{fig_MostFrequencyWords}.
Then, we separately count the popular words in the discussion of the COVID-19 vaccine in different countries.

\begin{figure*}
     \centering
     \subfloat[Prevalent words in tweets from India. 
     \label{IndiaWordCloud}] {\includegraphics[width=0.48\textwidth]{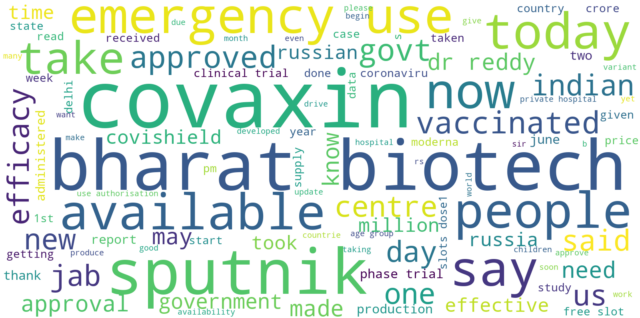} } \hfill
     \subfloat[Prevalent words in tweets from the United States.
     \label{AmericaWordCloud}]{\includegraphics[width=0.48\textwidth]{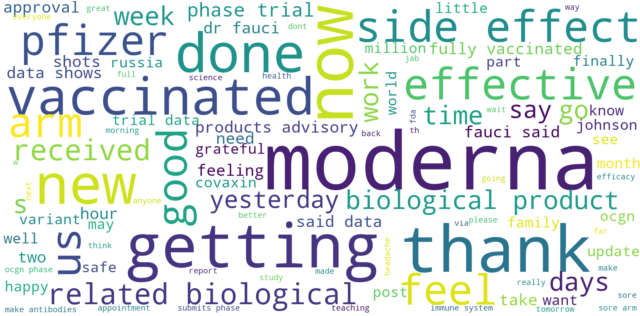} } \hfill \\
     \subfloat[Prevalent words in tweets from Canada.
     \label{CanadaWordCloud}]{\includegraphics[width=0.48\textwidth]{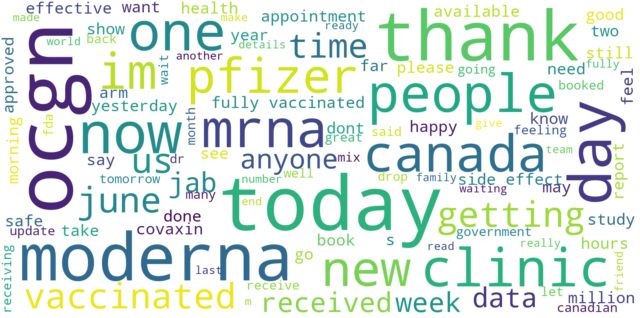}} \hfill  
    \subfloat[Prevalent words in tweets from the United Kingdom.
    \label{EnglandWordCloud}]{\includegraphics[width=0.48\textwidth]{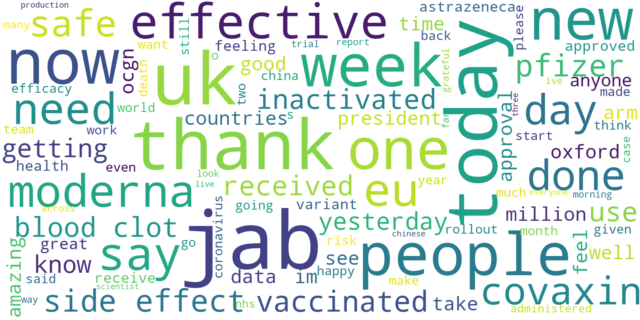}}\hfill

     \caption{Prevalent words in tweets from four countries in the dataset.}
     \label{fig_PrevalentWords}
\end{figure*}

We extract prevalent words from tweets in India, the United States, Canada, the United Kingdom and then use the word cloud to visualize them, as shown in Fig.~\ref{fig_PrevalentWords}.
According to Table~\ref{tab_ApprovedVaccine}, we learn that India has approved 4 of the 7 vaccines, the United States has approved 2 of the 7 vaccines, Canada and the United Kingdom have approved 3 of the 7 vaccines. 
Fig.~\ref{IndiaWordCloud} clearly shows that Indian people pay more attention to Bharat biotech (Covaxin) and Sputnik than Moderna and Oxford/AstraZeneca.
In the United States, Canada and the United Kingdom, Moderna and Pfizer are the most mentioned vaccines by users.
The word ``thank" is clearly visible in the word cloud, showing a positive attitude, as shown in Fig. \ref{AmericaWordCloud}, \ref{CanadaWordCloud} and \ref{EnglandWordCloud}.

\subsection{Prevalent Emotional Words by Vaccines}
\label{sec_EmotionalWordsByVaccine}
In this section, we pay attention to the high-frequency emotional vocabularies related to vaccines and gain a general understanding of people's attitudes toward different vaccines. 
We employ the VADER dictionary to filter the emotional words in the tweets, select the top 30 high-frequency emotional words for each vaccine, and then separate the words by polarity.
The results are shown in Table~\ref{tab_VaccinePositiveWordFrequency} and Table~\ref{tab_VaccineNegativeWordFrequency}, and we can see that the number of positive words is much higher than the number of negative words, such as ``thank'', ``approved'', ``effective'', ``safety'', ``hope''.
These words represent a positive attitude towards vaccines, trusting vaccines can protect us from infection.
We did not list neutral words because only two neutral words were mentioned in all vaccines' top 30 high-frequency emotional words.
\begin{table*}[h]
\centering
\caption{High-frequency positive words of different vaccines in tweets.}
\begin{tabular}{{ p{3.9cm}|p{7.5cm} }}
\toprule
\toprule
\textbf{Vaccine} & \textbf{Positive Vocabulary} \\[4pt] \midrule
\textbf{Pfizer/BioNTech}                                        & \begin{tabular}[c]{@{}l@{}}effective like thank good thanks approved great\\ feeling want safe heart grateful well better happy \\ protect please best ready hope approves share\end{tabular} \\ [13pt]\hline
\textbf{Sinopharm}                                              & \begin{tabular}[c]{@{}l@{}}well boost want special great ready approved \\ effective good  free approval thank like approves\\safe help please thanks  better positive support \\ number feeling gift best\end{tabular}  \\ [13pt]\hline
\textbf{Sinovac}                                                & \begin{tabular}[c]{@{}l@{}}approved validated approves reaches best feeling \\  launched like better effective thank approval good\\safe thanks well want validates number special\\please successfully boost\end{tabular} 
 \\ [13pt]\hline
\textbf{Oxford/AstraZeneca}                                     & \begin{tabular}[c]{@{}l@{}}proud feeling effective safety good safe pleased\\ happy great thank like hope delighted approved\\thanks well grateful want fine amazing please\end{tabular}  \\ [13pt]\hline
\textbf{Moderna}                                                & \begin{tabular}[c]{@{}l@{}}feeling ready thanks grateful great better like\\ best thank safe approval hope please good \\effective happy number free want approved well\\excited super help\end{tabular} \\ [13pt]\hline
\textbf{Covaxin}                                & \begin{tabular}[c]{@{}l@{}}safe best dear well want help trust effectively\\ positive good approval better like thanks\\effective great immune proud please approved \\thank free hope top\end{tabular} \\ [13pt]\hline
\textbf{Sputnik V(Gamaleya)}                                    & \begin{tabular}[c]{@{}l@{}}supreme number approved well help ready thank\\ thanks free allow best trust good effective like\\want approval great top launched approves please\\agreed\end{tabular}  \\ [13pt] \bottomrule \bottomrule
\end{tabular}
\label{tab_VaccinePositiveWordFrequency}
\end{table*}

\begin{table*}[h]
\centering
\caption{High-frequency negative words of different vaccines in tweets.}
\renewcommand{\arraystretch}{1.5}
\begin{tabular}{l|l}
\toprule
\toprule
\textbf{Vaccine} & \textbf{Negative Vocabulary}\\ \midrule
\textbf{Pfizer/BioNTech}                                        & no emergency risk warning death refused  \\ \hline
\textbf{Sinopharm}                                              & emergency low no missed    \\ \hline
\textbf{Sinovac}                                                & no low emergency death died  \\ \hline
\textbf{Oxford/AstraZeneca}                                     & no stop risk suspend sore rejected ill  \\ \hline
\textbf{Moderna}                                                & no ill sore emergency pain                \\ \hline
\textbf{Covaxin}                                                & severe strain emergency no shortage \\ \hline
\textbf{Sputnik V(Gamaleya)}                                    &  no emergency demand death fight \\  \bottomrule \bottomrule
\end{tabular}
\label{tab_VaccineNegativeWordFrequency}
\end{table*}

\subsection{Sentiment Analysis of Tweets}
\label{sec_SentimentAnalysis}

\begin{figure*}[!htbp]
     \centering
     \subfloat[India \label{IndiaSA}]{\includegraphics[width=0.49\textwidth]{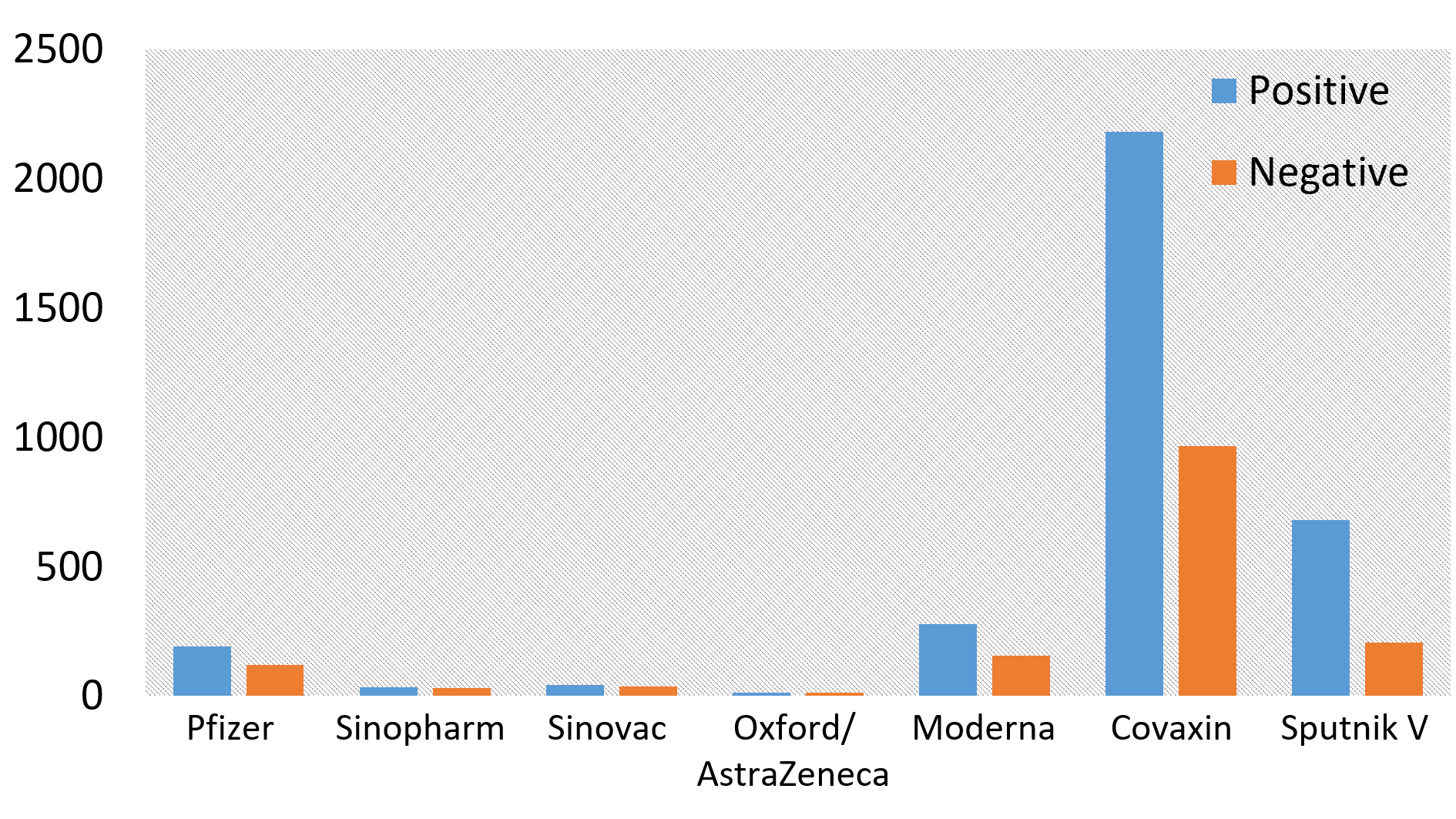}}\hfill
     \subfloat[the United States \label{United StatesSA}]{\includegraphics[width=0.49\textwidth]{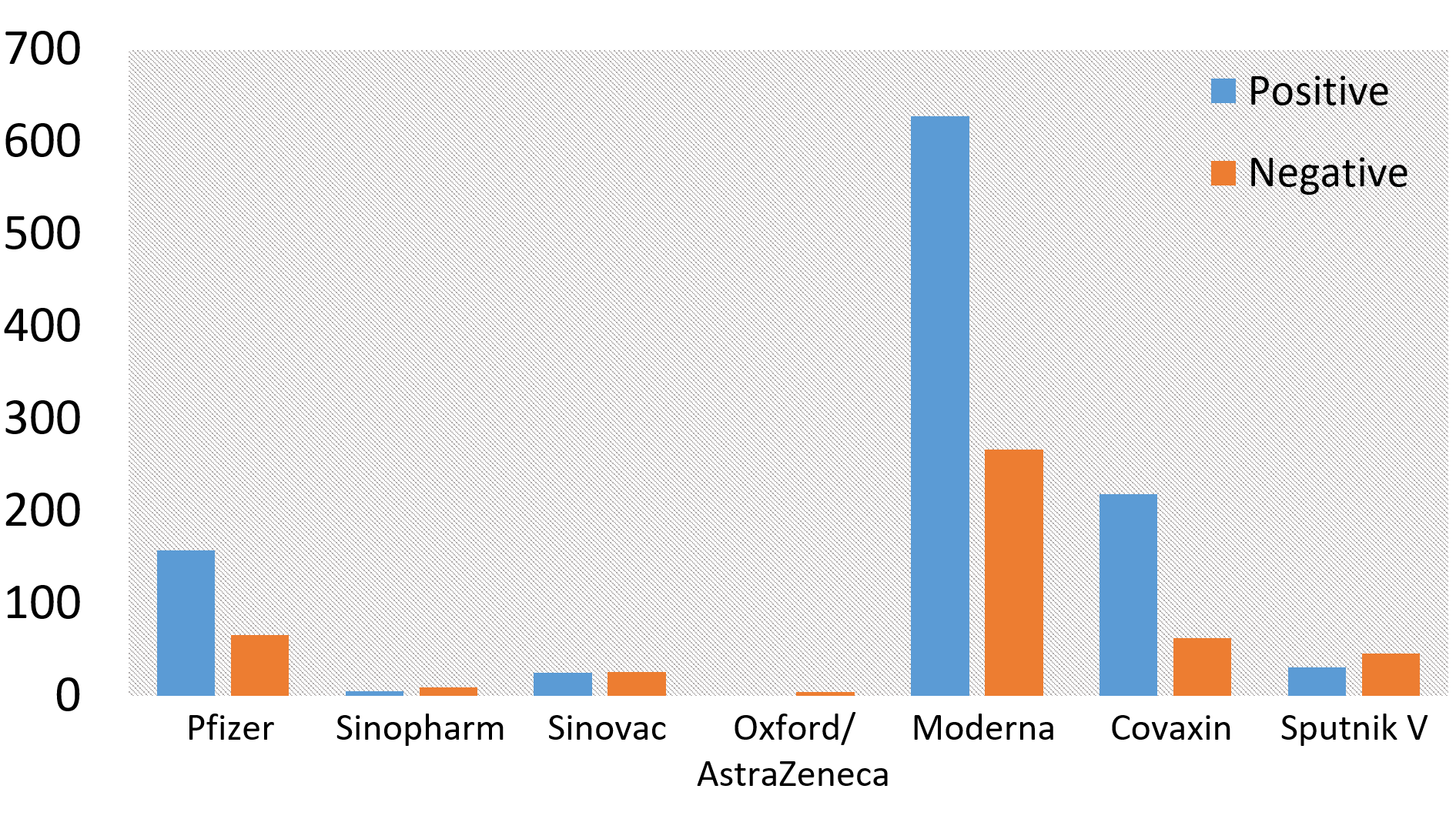}}\hfill \\
     \subfloat[Canada \label{CanadaSA}]{\includegraphics[width=0.49\textwidth]{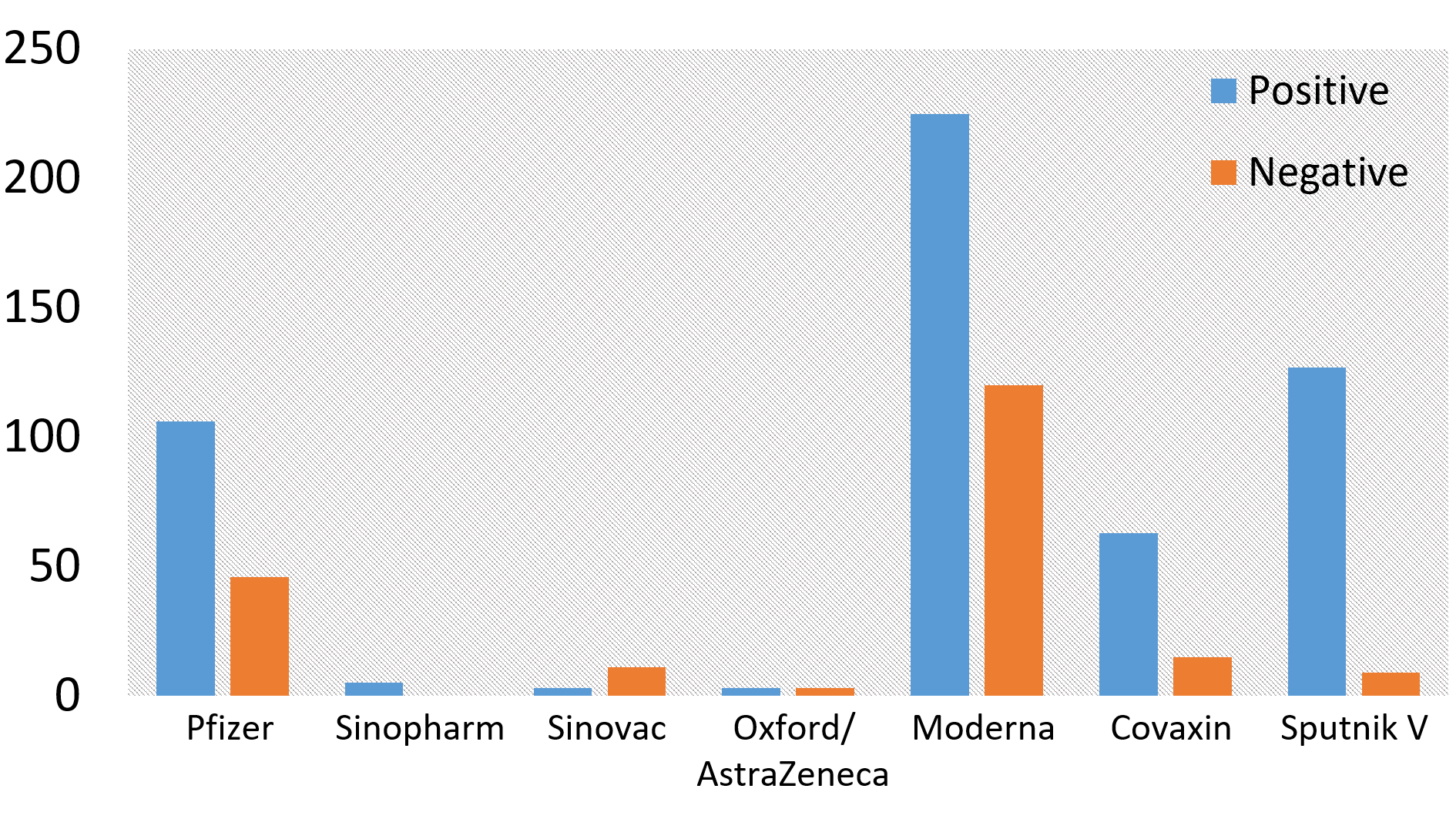}}\hfill 
     \subfloat[the United Kingdom \label{United KingdomSA}]{\includegraphics[width=0.49\textwidth]{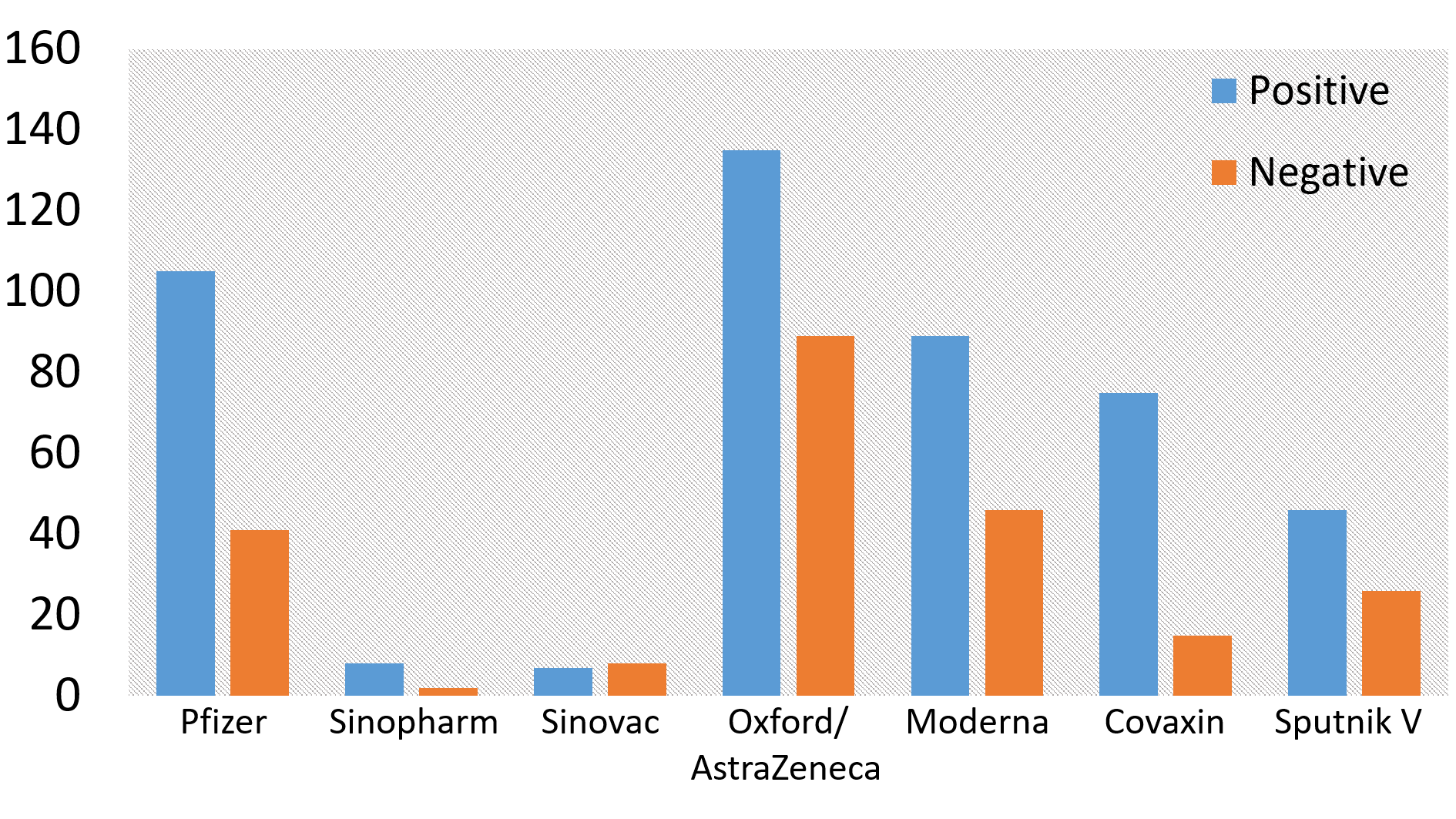} }\hfill \\
     \caption{Sentiment analysis of tweets for different vaccine brands in India, the United States, Canada and the United States}.
     \label{fig_CountrySA}
\end{figure*}

The sentiment polarity of each tweet is generated using the VADER tool as described previously.
Fig.~\ref{fig_CountrySA} presents the overall emotional distribution of tweets across the four countries to the seven vaccines over the study period.
Obviously, it can be seen that the number of positive tweets is greater than that of negative tweets, regardless of the brand, which shows that the majority of Twitter users maintain a positive attitude towards the vaccines.
According to Table \ref{tab_VaccinePositiveWordFrequency}, most of the positive tweets focused on the following aspects, such as believing that vaccines can provide effective protection, expecting that the vaccine will be approved and promoted as soon as possible, thanking the injection of the vaccine.
In contrast, negative tweets are mostly related to vaccine shortages, side effects after vaccination, and reports of deaths due to vaccination.

\begin{table}[]
\centering
\caption{The most discussed topics in positive tweets about the COVID-19 vaccine on Twitter during the study period. 
} 
\renewcommand{\arraystretch}{1.3}
\label{tab_PositiveTopics}
\begin{tabular}{{ p{1.8cm}|p{1.8cm}|p{1.8cm}|p{1.8cm}|p{1.8cm} }}
\toprule
\toprule
Topic 1 &Topic  2   &Topic  3            &Topic  4                  & Topic 5                    \\ \midrule
vaccine                                          & feel       & second      & go        & get        \\
say                                              & today      & shoot       & people    & be         \\
well                                             & day        & shot        & make      & still      \\
able                                             & good       & thank       & see       & happy      \\
friend                                           & week       & amp         & safe      & stay          \\
come                                             & time       & arm         & vaccinate & last       \\
many                                             & great      & do          & may       & vaccinated \\
free                                             & yesterday  & vaccination & work      & look       \\
soon                                             & thing      & hit         & let       & tell       \\
start                                            & back       & ready       & help      & lot            \\
\bottomrule
\bottomrule
\end{tabular}
\end{table}

\begin{table}[]
\centering
\caption{The most discussed topics in negative tweets about the COVID-19 vaccine on Twitter during the study period.
} 
\label{tab_NegativeTopics}
\renewcommand{\arraystretch}{1.3}
\begin{tabular}{{ p{1.8cm}|p{1.8cm}|p{1.8cm}|p{1.8cm}|p{1.8cm} }}
\toprule
\toprule
Topic 1 & Topic 2  & Topic 3          &Topic 4                    & Topic 5          \\ \midrule
vaccine                                          & arm        & be        & get    & take        \\
first                                            & dose       & shot      & week   & people      \\
day                                              & sore       & go        & ill    & know        \\
feel                                             & second     & amp       & report & stay           \\
still                                            & shoot      & make      & find   & home        \\
pain                                             & yesterday  & may       & kill   & vaccination \\
fever                                            & have       & update    & hear   & much        \\
little                                           & time       & come      & state  & good        \\
tired                                            & body       & think     & severe & month       \\
die                                              & tell       & vaccinate & expect & will   \\
\bottomrule
\bottomrule
\end{tabular}
\end{table}

\subsection{Topic Modeling of Tweets}
As mentioned in section~\ref{TopicmodelingOfTweets}, we use the coherence score to determine the optimal number of topics for topic modeling is 11.
In this section, we use LDA to generate the topics of the tweets to understand which aspects users concern about in the positive and negative tweets, respectively.
We count the number of tweets corresponding to different topics in positive tweets and negative tweets separately.
According to the popularity, the top 5 topics  discussed in positive and negative tweets are listed in Table~\ref{tab_PositiveTopics} and Table~\ref{tab_NegativeTopics}, where the most contributing words related to the topic are shown below the topic in the Tables. 
We get the consistent conclusions as in section~\ref{sec_EmotionalWordsByVaccine} and section~\ref{sec_SentimentAnalysis}.
In the positive tweets, people were grateful for being vaccinated in anticipation of returning to normal life; in negative tweets, most of them complain about side effects after vaccination, such as fever, sore arm, etc.
\section{Conclusion}
\label{sec_conclusions}
This study conducted a comprehensive analysis of COVID-19 vaccine-related tweets collected from Twitter between December 12, 2020 and July 2, 2021. A total of 75,665 tweets were used for this study, including seven vaccine brands, e.g., Pfizer/BioNTech, Sinopharm, Sinovac, Moderna, Oxford/AstraZeneca, Covaxin and Sputnik V.
According to statistics based on the location of tweet users, these tweets are mainly from India, the United States, Canada, and the United Kingdoms.
We first performed an overall analysis of the whole dataset and then a specific analysis of the four countries.
The sentiment analysis results showed that the overall sentiment polarity is positive, and the number of positive tweets is approximate twice the number of negative tweets.
When we drilled into country-level, it was found that the sentiment polarity scores of each country for the approved vaccines were consistent with the overall sentiment polarity scores.
But when it came to other vaccine brands, the number of negative tweets for some vaccines is higher than positive tweets, such as Sputnik V in the United States and Sinovac in Canada and the United States.
In the positive tweets, people expressed their gratitude for being able to be vaccinated.
They hope that with the help of the vaccination, the pandemic can be controlled as soon as possible and normal life can be resumed.
People mostly complained about side effects after vaccination in the negative tweets, such as fever, sore arm, etc.

In summary, this paper presented a case study of popular topics and sentiment analysis of tweets related to the COVID-19 vaccines. In the future, more interesting topics can be explored based on the current study.
For example, performing individual-level topic and sentiment analysis to help local communities locate the people that may suffer from the negative sentiments and thus take actions in advance.

\bibliographystyle{unsrt} 
\bibliography{refer}
\end{document}